# SECTORAL CONVERGENCE IN OUTPUT PER WORKER BETWEEN PORTUGUESE REGIONS

**Vitor João Pereira Domingues Martinho**

Unidade de I&D do Instituto Politécnico de Viseu
Av. Cor. José Maria Vale de Andrade
Campus Politécnico
3504 - 510 Viseu
**(PORTUGAL)**
**e-mail:** vdmartinho@esav.ipv.pt

**ABSTRACT**

The aim of this paper is to present a further contribution to the analysis of absolute convergence ( $\beta$ and $\sigma$ ), associated with the neoclassical theory, and conditional, associated with endogenous growth theory, of the sectoral productivity at regional level. Presenting some empirical evidence of absolute convergence of productivity for each of the economic sectors and industries in each of the regions of mainland Portugal (NUTS II and NUTS III) in the period 1986 to 1994 and from 1995 to 1999. The finest spatial unit NUTS III is only considered for each of the economic sectors in the period 1995 to 1999. They are also presented empirical evidence of conditional convergence of productivity, but only for each of the economic sectors of the NUTS II of Portugal, from 1995 to 1999. The structural variables used in the analysis of conditional convergence is the ratio of capital/output, the flow of goods/output and location ratio. The main conclusions should be noted that the signs of convergence are stronger in the first period than in the second and that convergence is conditional, especially in industry and in all sectors (1)(Martinho, 2011).

## 1. EMPIRICAL EVIDENCE OF ABSOLUTE CONVERGENCE, PANEL DATA

The purpose of this part of the work is to analyze the absolute convergence of output per worker (as a "proxy" of labor productivity), with the following equation ((2)Islam, 1995, based on the (3)Solow model, 1956):

$$\Delta \ln P_{it} = c + b \ln P_{i,t-1} + v_{it} \qquad (1)$$

Table 1 presents the results of absolute convergence of output per worker, obtained in the panel estimations for each of the economic sectors and the sectors to the total level of NUTS II, from 1986 to 1994 (a total of 45 observations, corresponding to regions 5 and 9 years).

The convergence results obtained in the estimations carried out are statistically satisfactory to each of the economic sectors and all sectors of the NUTS II.

**Table 1:** Analysis of convergence in productivity for each economic sectors of the five NUTS II of Portugal, for the period 1986 to 1994

| **Agriculture** | | | | | | | | | | | |
|---|---|---|---|---|---|---|---|---|---|---|---|
| **Method** | **Const.** | **$D_1$** | **$D_2$** | **$D_3$** | **$D_4$** | **$D_5$** | **Coef.** | **T.C.** | **DW** | **$R^2$** | **G.L.** |
| Pooling | 0.558 (1.200) | | | | | | -0.063 (-1.163) | -0.065 | 1.851 | 0.034 | 38 |
| LSDV | | 4.127* (4.119) | 4.207* (4.116) | 4.496* (4.121) | 4.636* (4.159) | 4.549* (4.091) | -0.514* (-4.108) | -0.722 | 2.202 | 0.352 | 34 |
| GLS | 0.357 (0.915) | | | | | | -0.040 (-0.871) | -0.041 | 1.823 | 0.020 | 38 |
| **Industry** | | | | | | | | | | | |
| **Method** | **Const.** | **$D_1$** | **$D_2$** | **$D_3$** | **$D_4$** | **$D_5$** | **Coef.** | **T.C.** | **DW** | **$R^2$** | **G.L.** |
| Pooling | 2.906* (2.538) | | | | | | -0.292* (-2.525) | -0.345 | 1.625 | 0.144 | 38 |
| LSDV | | 6.404* (4.345) | 6.459* (4.344) | 6.695* (4.341) | 6.986* (4.369) | 6.542* (4.334) | -0.667* (-4.344) | -1.100 | 1.679 | 0.359 | 34 |
| GLS | 3.260* (2.741) | | | | | | -0.328* (-2.729) | -0.397 | 1.613 | 0.164 | 38 |
| **Manufactured Industry** | | | | | | | | | | | |
| **Method** | **Const.** | **$D_1$** | **$D_2$** | **$D_3$** | **$D_4$** | **$D_5$** | **Coef.** | **T.C.** | **DW** | **$R^2$** | **G.L.** |
| Pooling | 1.806** (1.853) | | | | | | -0.186** (-1.845) | -0.206 | 1.935 | 0.082 | 38 |
| LSDV | | 6.625* (4.304) | 6.669* (4.303) | 6.941* (4.303) | 6.903* (4.318) | 6.626* (4.293) | -0.699* (-4.301) | -1.201 | 1.706 | 0.357 | 34 |
| GLS | 1.655** (1.753) | | | | | | -0.171** (-1.745) | -0.188 | 1.946 | 0.074 | 38 |
| **Services** | | | | | | | | | | | |
| **Method** | **Const.** | **$D_1$** | **$D_2$** | **$D_3$** | **$D_4$** | **$D_5$** | **Coef.** | **T.C.** | **DW** | **$R^2$** | **G.L.** |
| Pooling | 5.405* (4.499) | | | | | | -0.554* (-4.477) | -0.807 | 1.874 | 0.345 | 38 |
| LSDV | | 7.193* (5.290) | 7.169* (5.301) | 7.313* (5.284) | 7.153* (5.292) | 7.273* (5.293) | -0.741* (-5.275) | -1.351 | 2.051 | 0.451 | 34 |
| GLS | 5.627* (4.626) | | | | | | -0.577* (-4.604) | -0.860 | 1.886 | 0.358 | 38 |
| **Services (without public sector)** | | | | | | | | | | | |
| **Method** | **Const.** | **$D_1$** | **$D_2$** | **$D_3$** | **$D_4$** | **$D_5$** | **Coef.** | **T.C.** | **DW** | **$R^2$** | **G.L.** |
| Pooling | 5.865* (4.079) | | | | | | -0.589* (-4.073) | -0.889 | 1.679 | 0.304 | 38 |
| LSDV | | 6.526* (4.197) | 6.523* (4.195) | 6.635* (4.191) | 6.506* (4.176) | 6.561* (4.192) | -0.658* (-4.188) | -1.073 | 1.684 | 0.342 | 34 |
| GLS | 5.027* (3.656) | | | | | | -0.505* (-3.649) | -0.703 | 1.682 | 0.260 | 38 |
| **All sectors** | | | | | | | | | | | |
| **Method** | **Const.** | **$D_1$** | **$D_2$** | **$D_3$** | **$D_4$** | **$D_5$** | **Coef.** | **T.C.** | **DW** | **$R^2$** | **G.L.** |
| Pooling | 3.166* (3.603) | | | | | | -0.328* (-3.558) | -0.397 | 1.785 | 0.250 | 38 |
| LSDV | | 6.080* (5.361) | 6.030* (5.374) | 6.308* (5.347) | 6.202* (5.379) | 6.193* (5.359) | -0.643* (-5.333) | -1.030 | 2.181 | 0.460 | 34 |
| GLS | 3.655* (3.916) | | | | | | -0.379* (-3.874) | -0.476 | 1.815 | 0.283 | 38 |

**Note: Const. Constant; Coef., Coefficient, TC, annual rate of convergence; * Coefficient statistically significant at 5%, ** Coefficient statistically significant at 10%, GL, Degrees of freedom; LSDV, method of fixed effects with variables dummies; D1 ... D5, five variables dummies corresponding to five different regions, GLS, random effects method.**

Table 2 shows results also for each of the economic sectors and all sectors of the NUTS II of Portugal, but now for the period 1995 to 1999.

**Table 2:** Analysis of convergence in productivity for each of the sectors and in NUTS II of Portugal, for the period 1995 to 1999

| | | | | | | | | | | | |
|---|---|---|---|---|---|---|---|---|---|---|---|
| **Agriculture** | | | | | | | | | | | |
| **Method** | **Const.** | **$D_1$** | **$D_2$** | **$D_3$** | **$D_4$** | **$D_5$** | **Coef.** | **T.C.** | **DW** | **$R^2$** | **G.L.** |
| Pooling | -0.038 (-0.089) | | | | | | 0.005 (0.101) | 0.005 | 2.113 | 0.001 | 18 |
| LSDV | | 5.672* (2.662) | 5.703* (2.653) | 6.288* (2.674) | 6.403* (2.657) | 6.230* (2.692) | -0.673* (-2.666) | -1.118 | 2.048 | 0.423 | 14 |
| GLS | -0.132 (-0.438) | | | | | | 0.015 (0.456) | 0.015 | 1.867 | 0.011 | 18 |
| **Industry** | | | | | | | | | | | |
| **Method** | **Const.** | **$D_1$** | **$D_2$** | **$D_3$** | **$D_4$** | **$D_5$** | **Coef.** | **T.C.** | **DW** | **$R^2$** | **G.L.** |
| Pooling | 0.754** (1.991) | | | | | | -0.073** (-1.880) | -0.076 | 2.194 | 0.164 | 18 |
| LSDV | | 2.965* (2.328) | 3.018* (2.335) | 3.107* (2.330) | 3.089* (2.308) | 2.994* (2.348) | -0.306* (-2.297) | -0.365 | 2.377 | 0.397 | 14 |
| GLS | 0.640* (2.433) | | | | | | -0.061* (-2.273) | -0.063 | 2.032 | 0.223 | 18 |
| **Manufactured industry** | | | | | | | | | | | |
| **Method** | **Const.** | **$D_1$** | **$D_2$** | **$D_3$** | **$D_4$** | **$D_5$** | **Coef.** | **T.C.** | **DW** | **$R^2$** | **G.L.** |
| Pooling | 1.426* (2.249) | | | | | | -0.140* (-2.134) | -0.151 | 1.369 | 0.202 | 18 |
| LSDV | | 2.697* (2.291) | 2.761* (2.290) | 2.834* (2.281) | 2.808* (2.358) | 2.742* (2.339) | -0.279* (-2.251) | -0.327 | 1.978 | 0.444 | 14 |
| GLS | 1.502* (2.245) | | | | | | -0.148* (-2.135) | -0.160 | 1.429 | 0.202 | 18 |
| **Sercices** | | | | | | | | | | | |
| **Method** | **Const.** | **$D_1$** | **$D_2$** | **$D_3$** | **$D_4$** | **$D_5$** | **Coef.** | **T.C.** | **DW** | **$R^2$** | **G.L.** |
| Pooling | -0.058 (-0.180) | | | | | | 0.011 (0.333) | 0.011 | 2.282 | 0.006 | 18 |
| LSDV | | 0.957 (1.365) | 0.943 (1.355) | 0.986 (1.379) | 0.946 (1.369) | 0.971 (1.375) | -0.093 (-1.299) | -0.098 | 2.929 | 0.212 | 14 |
| GLS | -0.265 (-1.497) | | | | | | 0.032** (1.774) | 0.031 | 1.955 | 0.149 | 18 |
| **All sectors** | | | | | | | | | | | |
| **Method** | **Const.** | **$D_1$** | **$D_2$** | **$D_3$** | **$D_4$** | **$D_5$** | **Coef.** | **T.C.** | **DW** | **$R^2$** | **G.L.** |
| Pooling | -0.044 (-0.154) | | | | | | 0.009 (0.316) | 0.009 | 1.803 | 0.006 | 18 |

| | | | | | | | | | | |
|---|---|---|---|---|---|---|---|---|---|---|
| LSDV | | 0.954 (1.383) | 0.949 (1.380) | 0.999 (1.392) | 0.948 (1.356) | 0.980 (1.398) | -0.095 (-1.316) | -0.100 | 2.714 | 0.365 | 14 |
| GLS | 0.014 (0.045) | | | | | | 0.003 (0.100) | 0.003 | 1.925 | 0.001 | 18 |

Looking at the coefficient of convergence, we now find evidence of absolute convergence only for agriculture, industry and manufactured industry.

Are presented subsequently in Table 3 the results of the absolute convergence of output per worker, obtained in the panel estimations for each of the sectors and all sectors, now at the level of NUTS III during the period 1995 to 1999.

The results of convergence are statistically satisfactory all sectors and sectors to the total level of NUTS III.

**Table 3**: Analysis of convergence in productivity for each of the economic sectors at the level of NUTS III of Portugal, for the period 1995 to 1999

| **Agriculture** | | | | | | |
|---|---|---|---|---|---|---|
| **Method** | **Const.** | **Coef.** | **T.C.** | **DW** | **$R^2$** | **G.L.** |
| Pooling | 0.017 (0.086) | -0.003 (-0.146) | -0.003 | 2.348 | 0.000 | 110 |
| LSDV | | -0.938* (-9.041) | -2.781 | 2.279 | 0.529 | 83 |
| GLS | -0.219* (-3.633) | 0.024* (3.443) | 0.024 | 1.315 | 0.097 | 110 |
| **Industry** | | | | | | |
| **Method** | **Const.** | **Coef.** | **T.C.** | **DW** | **$R^2$** | **G.L.** |
| Pooling | 0.770* (4.200) | -0.076* (-4.017) | -0.079 | 1.899 | 0.128 | 110 |
| LSDV | | -0.511* (-7.784) | -0.715 | 2.555 | 0.608 | 83 |
| GLS | 0.875* (4.154) | -0.086* (-3.994) | -0.090 | 2.062 | 0.127 | 110 |
| **Services** | | | | | | |
| **Method** | **Const.** | **Coef.** | **T.C.** | **DW** | **$R^2$** | **G.L.** |
| Pooling | 0.258 (1.599) | -0.022 (-1.314) | -0.022 | 1.955 | 0.016 | 110 |
| LSDV | | -0.166* (-5.790) | -0.182 | 2.665 | 0.382 | 83 |
| GLS | 0.089 (0.632) | -0.004 (-0.303) | -0.004 | 1.868 | 0.001 | 110 |
| **All sectors** | | | | | | |
| **Method** | **Const.** | **Coef.** | **T.C.** | **DW** | **$R^2$** | **G.L.** |
| "Pooling" | 0.094 (0.833) | -0.005 (-0.445) | -0.005 | 2.234 | 0.002 | 110 |
| LSDV | | -0.156* (-3.419) | -0.170 | 2.664 | 0.311 | 83 |
| GLS | 0.079 (0.750) | -0.004 (-0.337) | -0.004 | 2.169 | 0.001 | 110 |

Table 4 presents the results for the absolute convergence of output per worker, in the estimations obtained for each of the manufactured industry of NUTS II, from 1986 to 1994.

The convergence results obtained are statistically satisfactory for all manufacturing industries of NUTS II.

**Table 4**: Analysis of convergence in productivity for each of the manufacturing industries at the five NUTS II of Portugal, for the period 1986 to 1994

| **Metals industry** | | | | | | | | | | | |
|---|---|---|---|---|---|---|---|---|---|---|---|
| **Method** | **Const.** | **$D_1$** | **$D_2$** | **$D_3$** | **$D_4$** | **$D_5$** | **Coef.** | **T.C.** | **DW** | **$R^2$** | **G.L.** |
| Pooling | 0.190 (0.190) | | | | | | -0.024 (-0.241) | -0.024 | 1.646 | 0.002 | 30 |
| LSDV | | 2.171** (1.769) | 2.143** (1.753) | 2.161** (1.733) | 2.752** (1.988) | --- | -0.239** (-1.869) | -0.273 | 1.759 | 0.198 | 27 |
| GLS | 0.407 (0.394) | | | | | | -0.046 (-0.445) | -0.047 | 1.650 | 0.007 | 30 |
| **MInerals industry** | | | | | | | | | | | |
| **Method** | **Const.** | **$D_1$** | **$D_2$** | **$D_3$** | **$D_4$** | **$D_5$** | **Coef.** | **T.C.** | **DW** | **$R^2$** | **G.L.** |
| Pooling | 0.738 (0.903) | | | | | | -0.085 (-0.989) | -0.089 | 1.935 | 0.025 | 38 |
| LSDV | | 1.884* (2.051) | 1.970* (2.112) | 2.004* (2.104) | 1.926* (2.042) | 1.731** (1.930) | -0.208* (-2.129) | -0.233 | 2.172 | 0.189 | 34 |
| GLS | 0.967<br>(1.162) | | | | | | -0.109<br>(-1.246) | -0.115 | 1.966 | 0.039 | 38 |
| **Chemical industry** | | | | | | | | | | | |
| **Method** | **Const.** | **$D_1$** | **$D_2$** | **$D_3$** | **$D_4$** | **$D_5$** | **Coef.** | **T.C.** | **DW** | **$R^2$** | **G.L.** |
| Pooling | 2.312**<br>(1.992) | | | | | | -0.225**<br>(-1.984) | -0.255 | 2.017 | 0.104 | 34 |
| LSDV | | 6.104*<br>(3.750) | 6.348*<br>(3.778) | 6.381*<br>(3.774) | 6.664*<br>(3.778) | 6.254*<br>(3.777) | -0.621*<br>(-3.769) | -0.970 | 1.959 | 0.325 | 30 |
| GLS | 2.038**<br>(1.836) | | | | | | -0.198**<br>(-1.826) | -0.221 | 2.034 | 0.089 | 34 |
| **Electric goods industry** | | | | | | | | | | | |
| **Method** | **Const.** | **$D_1$** | **$D_2$** | **$D_3$** | **$D_4$** | **$D_5$** | **Coef.** | **T.C.** | **DW** | **$R^2$** | **G.L.** |
| Pooling | 0.781<br>(0.789) | | | | | | -0.083<br>(-0.784) | -0.087 | 1.403 | 0.016 | 38 |
| LSDV | | 3.634*<br>(2.363) | 3.552*<br>(2.360) | 3.673*<br>(2.362) | 3.636*<br>(2.376) | 3.429*<br>(2.324) | -0.381*<br>(-2.355) | -0.480 | 1.259 | 0.167 | 34 |
| GLS | 0.242<br>(0.285) | | | | | | -0.025<br>(-0.279) | -0.025 | 1.438 | 0.002 | 38 |
| **Transport equipments industry** | | | | | | | | | | | |
| **Method** | **Const.** | **$D_1$** | **$D_2$** | **$D_3$** | **$D_4$** | **$D_5$** | **Coef.** | **T.C.** | **DW** | **$R^2$** | **G.L.** |
| Pooling | 4.460*<br>(3.110) | | | | | | -0.464*<br>(-3.136) | -0.624 | 2.258 | 0.206 | 38 |
| LSDV | | 8.061*<br>(4.948) | 8.526*<br>(5.007) | 8.614*<br>(4.986) | 8.696*<br>(4.998) | 8.077*<br>(4.961) | -0.871*<br>(-5.014) | -2.048 | 2.049 | 0.429 | 34 |
| GLS | 5.735*<br>(3.780) | | | | | | -0.596*<br>(-3.807) | -0.906 | 2.159 | 0.276 | 38 |
| **Food industry** | | | | | | | | | | | |
| **Method** | **Const.** | **$D_1$** | **$D_2$** | **$D_3$** | **$D_4$** | **$D_5$** | **Coef.** | **T.C.** | **DW** | **$R^2$** | **G.L.** |
| Pooling | 0.314<br>(0.515) | | | | | | -0.027<br>(-0.443) | -0.027 | 1.858 | 0.005 | 38 |
| LSDV | | 2.841*<br>(2.555) | 2.777*<br>(2.525) | 2.899*<br>(2.508) | 2.617*<br>(2.471) | 2.593*<br>(2.470) | -0.274*<br>(-2.469) | -0.320 | 1.786 | 0.198 | 34 |
| GLS | 0.090<br>(0.166) | | | | | | -0.005<br>(-0.085) | -0.005 | 1.851 | 0.001 | 38 |
| **Textile industry** | | | | | | | | | | | |
| **Method** | **Const.** | **$D_1$** | **$D_2$** | **$D_3$** | **$D_4$** | **$D_5$** | **Coef.** | **T.C.** | **DW** | **$R^2$** | **G.L.** |
| Pooling | 4.276*<br>(4.639) | | | | | | -0.462*<br>(-4.645) | -0.620 | 1.836 | 0.388 | 34 |
| LSDV | | 5.556*<br>(4.288) | 5.487*<br>(4.276) | 5.506*<br>(4.272) | 5.561*<br>(4.253) | 5.350*<br>(4.431) | -0.595*<br>(-4.298) | -0.904 | 1.816 | 0.431 | 30 |
| GLS | 3.212*<br>(6.336) | | | | | | -0.347*<br>(-6.344) | -0.426 | 1.848 | 0.542 | 34 |
| **Paper industry** | | | | | | | | | | | |
| **Method** | **Const.** | **$D_1$** | **$D_2$** | **$D_3$** | **$D_4$** | **$D_5$** | **Coef.** | **T.C.** | **DW** | **$R^2$** | **G.L.** |
| Pooling | 2.625*<br>(2.332) | | | | | | -0.271*<br>(-2.366) | -0.316 | 1.534 | 0.128 | 38 |
| LSDV | | 3.703*<br>(2.803) | 3.847*<br>(2.840) | 3.837*<br>(2.813) | 3.684*<br>(2.812) | 3.521*<br>(2.782) | -0.382*<br>(-2.852) | -0.481 | 1.516 | 0.196 | 34 |
| GLS | 1.939**<br>(1.888) | | | | | | -0.201**<br>(-1.924) | -0.224 | 1.556 | 0.089 | 38 |
| **Several industry** | | | | | | | | | | | |
| **Method** | **Const.** | **$D_1$** | **$D_2$** | **$D_3$** | **$D_4$** | **$D_5$** | **Coef.** | **T.C.** | **DW** | **$R^2$** | **G.L.** |
| Pooling | 5.518*<br>(4.004) | | | | | | -0.605*<br>(-4.004) | -0.929 | 2.121 | 0.297 | 38 |
| LSDV | | 7.802*<br>(5.036) | 7.719*<br>(5.022) | 7.876*<br>(5.033) | 7.548*<br>(5.023) | 7.660*<br>(5.018) | -0.847*<br>(-5.032) | -1.877 | 2.024 | 0.428 | 34 |
| GLS | 6.053*<br>(4.308) | | | | | | -0.664*<br>(-4.309) | -1.091 | 2.081 | 0.328 | 38 |

Table 5 shows results also for each of the manufacturing industries of the NUTS II of Portugal, but now for the period 1995 to 1999.

**Table 5**: Analysis of convergence in productivity for each of the manufacturing industries at the five NUTS II of Portugal, for the period 1995 to 1999

| **Metals industry** | | | | | | | | | | | |
|---|---|---|---|---|---|---|---|---|---|---|---|
| **Method** | **Const.** | **$D_1$** | **$D_2$** | **$D_3$** | **$D_4$** | **$D_5$** | **Coef.** | **T.C.** | **DW** | **$R^2$** | **G.L.** |
| Pooling | 1.108*<br>(3.591) | | | | | | -0.111*<br>(-3.353) | -0.118 | 2.457 | 0.384 | 18 |
| LSDV | | 1.476<br>(1.143) | 1.496<br>(1.183) | 1.503<br>(1.129) | 1.451<br>(1.186) | 1.459<br>(1.233) | -0.151<br>(-1.115) | -0.164 | 2.424 | 0.416 | 14 |
| GLS | 1.084* (7.366) | | | | | | -0.108* (-6.866) | -0.114 | 2.176 | 0.724 | 18 |
| **Minerals industry** | | | | | | | | | | | |
| **Method** | **Const.** | **$D_1$** | **$D_2$** | **$D_3$** | **$D_4$** | **$D_5$** | **Coef.** | **T.C.** | **DW** | **$R^2$** | **G.L.** |
| Pooling | -0.455 (-1.236) | | | | | | 0.052 (1.409) | 0.051 | 1.601 | 0.099 | 18 |
| LSDV | | 2.158* (2.222) | 2.280* (2.265) | 2.287* (2.227) | 2.194* (2.248) | 2.417* (2.306) | -0.221* (-2.192) | -0.250 | 1.359 | 0.567 | 14 |
| GLS | -0.356 (-0.854) | | | | | | 0.042 (1.007) | 0.041 | 1.628 | 0.053 | 18 |
| **Chemical industry** | | | | | | | | | | | |
| **Method** | **Const.** | **$D_1$** | **$D_2$** | **$D_3$** | **$D_4$** | **$D_5$** | **Coef.** | **T.C.** | **DW** | **$R^2$** | **G.L.** |
| Pooling | 1.236 (1.026) | | | | | | -0.115 (-0.966) | -0.122 | 1.049 | 0.049 | 18 |
| LSDV | | 5.320* (4.493) | 5.281* (4.452) | 5.447* (4.449) | 5.858* (4.711) | 5.072* (4.501) | -0.525* (-4.470) | -0.744 | 2.432 | 0.702 | 14 |
| GLS | 3.136* (2.532) | | | | | | -0.302* (-2.477) | -0.360 | 1.174 | 0.254 | 18 |
| **Electric goods industry** | | | | | | | | | | | |
| **Method** | **Const.** | **$D_1$** | **$D_2$** | **$D_3$** | **$D_4$** | **$D_5$** | **Coef.** | **T.C.** | **DW** | **$R^2$** | **G.L.** |
| Pooling | 1.936 (1.289) | | | | | | -0.196 (-1.271) | -0.218 | 1.945 | 0.082 | 18 |
| LSDV | | 4.729 (1.504) | 4.775 (1.507) | 4.818 (1.490) | 4.590 (1.463) | 4.671 (1.519) | -0.482 (-1.488) | -0.658 | 2.038 | 0.342 | 14 |
| GLS | 2.075 (1.299) | | | | | | -0.211 (-1.283) | -0.237 | 1.976 | 0.084 | 18 |
| **Transport equipments industry** | | | | | | | | | | | |
| **Method** | **Const.** | **$D_1$** | **$D_2$** | **$D_3$** | **$D_4$** | **$D_5$** | **Coef.** | **T.C.** | **DW** | **$R^2$** | **G.L.** |
| Pooling | 2.429* (2.264) | | | | | | -0.237* (-2.179) | -0.270 | 1.837 | 0.209 | 18 |
| LSDV | | 8.626* (10.922) | 8.647* (10.973) | 9.051* (10.924) | 8.537* (10.917) | 8.356* (10.866) | -0.867* (-10.811) | -2.017 | 2.000 | 0.896 | 14 |
| GLS | 3.507* (3.025) | | | | | | -0.346* (-2.947) | -0.425 | 1.649 | 0.326 | 18 |
| **Food industry** | | | | | | | | | | | |
| **Method** | **Const.** | **$D_1$** | **$D_2$** | **$D_3$** | **$D_4$** | **$D_5$** | **Coef.** | **T.C.** | **DW** | **$R^2$** | **G.L.** |
| Pooling | 0.873 (1.619) | | | | | | -0.082 (-1.453) | -0.086 | 2.921 | 0.105 | 18 |
| LSDV | | -0.516 (-0.300) | -0.521 (-0.308) | -0.532 (-0.304) | -0.425 (-0.259) | -0.435 (-0.268) | 0.060 (0.341) | 0.058 | 2.230 | 0.208 | 14 |
| GLS | 1.027* (4.163) | | | | | | -0.098* (-3.800) | -0.103 | 2.251 | 0.445 | 18 |
| **Textile industry** | | | | | | | | | | | |
| **Method** | **Const.** | **$D_1$** | **$D_2$** | **$D_3$** | **$D_4$** | **$D_5$** | **Coef.** | **T.C.** | **DW** | **$R^2$** | **G.L.** |
| Pooling | 0.788** (2.048) | | | | | | -0.080** (-1.882) | -0.083 | 1.902 | 0.165 | 18 |
| LSDV | | 0.514 (0.261) | 0.525 (0.270) | 0.515 (0.262) | 0.522 (0.272) | 0.541 (0.301) | -0.051 (-0.239) | -0.052 | 1.919 | 0.167 | 14 |
| GLS | 0.802* (20.052) | | | | | | -0.081* (-18.461) | -0.085 | 1.719 | 0.950 | 18 |
| **Paper industry** | | | | | | | | | | | |
| **Method** | **Const.** | **$D_1$** | **$D_2$** | **$D_3$** | **$D_4$** | **$D_5$** | **Coef.** | **T.C.** | **DW** | **$R^2$** | **G.L.** |
| Pooling | 0.735 (1.524) | | | | | | -0.073 (-1.471) | -0.076 | 2.341 | 0.107 | 18 |
| LSDV | | 5.201 (1.479) | 5.454 (1.462) | 5.410 (1.467) | 5.053 (1.470) | 4.970 (1.486) | -0.533 (-1.465) | -0.761 | 1.939 | 0.227 | 14 |
| GLS | 0.654* (3.329) | | | | | | -0.064* (-3.198) | -0.066 | 2.185 | 0.362 | 18 |
| **Several industry** | | | | | | | | | | | |
| **Method** | **Const.** | **$D_1$** | **$D_2$** | **$D_3$** | **$D_4$** | **$D_5$** | **Coef.** | **T.C.** | **DW** | **$R^2$** | **G.L.** |
| Pooling | -0.338 (-0.463) | | | | | | 0.042 (0.531) | 0.041 | 2.651 | 0.015 | 18 |
| LSDV | | 3.734** (1.949) | 3.883** (1.962) | 3.940** (1.966) | 3.817** (1.967) | 3.647** (1.934) | -0.402** (-1.930) | -0.514 | 2.905 | 0.303 | 14 |
| GLS | -0.904* (-3.791) | | | | | | 0.102* (4.003) | 0.097 | 1.922 | 0.471 | 18 |

## 2. EMPIRICAL EVIDENCE OF CONDITIONAL CONVERGENCE WITH PANEL DATA

This part of the work aims to analyze the conditional convergence of labor productivity sectors (using as a "proxy" output per worker) between the different NUTS II of Portugal, from 1995 to 1999.

Given these limitations and the availability of data, it was estimated in this part of the work equation (1) introducing some structural variables, namely, the ratio of gross fixed capital/output (such as "proxy" for the accumulation of capital/output ), the flow ratio of goods/output (as a "proxy" for transport costs) and the location quotient (calculated as the ratio between the number of regional employees in a given sector and the number of national employees in this sector on the ratio between the number regional employment and the number of national employees) ((4) Sala-i-Martin, 1996).

Considering the results obtained and presented in Table 6 (for conditional convergence), compared with those presented in Table 2 (absolute convergence), it appears that only in industry and all sectors is that the coefficient of convergence improve.

**Table 6**: Analysis of conditional convergence in productivity for each of the sectors at NUTS II of Portugal, for the period 1995 to 1999

**Agriculture**

| Method | Const. | $D_1$ | $D_2$ | $D_3$ | $D_4$ | $D_5$ | Coef.1 | Coef.2 | Coef.3 | Coef.4 | DW | $R^2$ | G.L. |
|---|---|---|---|---|---|---|---|---|---|---|---|---|---|
| Pooling | 0.114 (0.247) | | | | | | -0.020 (-0.392) | 0.388 (0.592) | 0.062 (1.267) | -0.062 (-1.160) | 2.527 | 0.136 | 15 |
| LSDV | | 5.711* (2.333) | 5.856* (2.385) | 6.275* (2.299) | 6.580* (2.383) | 6.517* (2.431) | -0.649* (-2.248) | -0.134 (-0.134) | -0.132 (-0.437) | -0.102 (-0.189) | 2.202 | 0.469 | 11 |
| GLS | -0.020 (-0.221) | | | | | | -0.004 (-0.416) | 0.284 (1.419) | 0.059* (4.744) | -0.053* (-4.163) | 2.512 | 0.797 | 15 |

**Industry**

| Method | Const. | $D_1$ | $D_2$ | $D_3$ | $D_4$ | $D_5$ | Coef.1 | Coef.2 | Coef.3 | Coef.5 | DW | $R^2$ | G.L. |
|---|---|---|---|---|---|---|---|---|---|---|---|---|---|
| Pooling | 3.698* (4.911) | | | | | | -0.336* (-5.055) | 0.269* (3.229) | -0.125* (-3.888) | -0.297* (-3.850) | 2.506 | 0.711 | 15 |
| LSDV | | 4.486* (6.153) | 4.386* (6.700) | 4.435* (7.033) | 4.335* (6.967) | 4.111* (6.977) | -0.421* (-6.615) | 0.530* (6.222) | 0.018 (0.412) | -0.397 (-0.854) | 2.840 | 0.907 | 11 |
| GLS | 3.646* (4.990) | | | | | | -0.332* (-5.144) | 0.279* (3.397) | -0.123* (-3.899) | -0.290* (-3.828) | 2.597 | 0.719 | 15 |

**Manufactured industry**

| Method | Const. | $D_1$ | $D_2$ | $D_3$ | $D_4$ | $D_5$ | Coef.1 | Coef.2 | Coef.3 | Coef.6 | DW | $R^2$ | G.L. |
|---|---|---|---|---|---|---|---|---|---|---|---|---|---|
| Pooling | 0.468 (0.690) | | | | | | -0.053 (-0.870) | 0.285* (4.502) | 0.013 (0.359) | 0.010 (0.167) | 2.177 | 0.804 | 15 |
| LSDV | | 2.850** (2.065) | 2.461** (2.081) | 2.068** (2.067) | 1.851** (2.022) | 1.738* (2.172) | -0.123 (-1.772) | 0.296* (5.185) | -0.097 (-1.448) | -1.119 (-1.787) | 1.770 | 0.923 | 11 |
| GLS | 0.513 (0.729) | | | | | | -0.057 (-0.906) | 0.289* (4.539) | 0.009 (0.252) | 0.008 (0.123) | 2.169 | 0.800 | 15 |

**Services**

| Method | Const. | $D_1$ | $D_2$ | $D_3$ | $D_4$ | $D_5$ | Coef.1 | Coef.2 | Coef.3 | Coef.7 | DW | $R^2$ | G.L. |
|---|---|---|---|---|---|---|---|---|---|---|---|---|---|
| Pooling | 0.472 (1.209) | | | | | | -0.046 (-1.110) | -0.118 (-1.653) | -0.013 (-1.401) | 0.081** (2.071) | 2.367 | 0.268 | 15 |
| LSDV | | 1.774 (1.329) | 1.831 (1.331) | 2.140 (1.324) | 1.955 (1.344) | 2.217 (1.345) | -0.109 (-1.160) | -0.137 (-1.400) | -0.075 (-1.380) | -0.698 (-1.024) | 2.393 | 0.399 | 11 |
| GLS | 0.238 (0.790) | | | | | | -0.022 (-0.718) | -0.079 (-0.967) | -0.008 (-1.338) | 0.060* (2.126) | 1.653 | 0.613 | 15 |

**All sectors**

| Method | Const. | $D_1$ | $D_2$ | $D_3$ | $D_4$ | $D_5$ | Coef.1 | Coef.2 | Coef.3 | Coef.4 | Coef.5 | Coef.7 | DW | $R^2$ | G.L. |
|---|---|---|---|---|---|---|---|---|---|---|---|---|---|---|---|
| Pooling | 0.938 (0.910) | | | | | | -0.077 (-1.04) | -0.152 (-0.88) | -0.011 (-0.71) | -0.029 (-0.28) | -0.057 (-0.20) | 0.005 (0.009) | 2.738 | 0.458 | 13 |
| LSDV | | -0.797 (-0.67) | -0.645 (-0.54) | -0.545 (-0.41) | -0.521 (-0.42) | -0.263 (-0.20) | 0.011 (0.130) | -0.483* (-2.72) | -0.155* (-2.79) | 0.085 (0.802) | 0.465 (1.279) | 0.344 (0.590) | 2.591 | 0.792 | 9 |
| GLS | 1.018 (0.976) | | | | | | -0.088 (-1.16) | -0.182 (-1.14) | -1.034 (-1.03) | -0.026 (-0.26) | -0.050 (-0.17) | 0.023 (0.043) | 2.676 | 0.854 | 13 |

**Note: Const. Constant; Coef1., Coefficient of convergence; Coef.2, Coefficient of the ratio capital/output; Coef.3, Coefficient of the ratio of flow goods/output; Coef.4, Coefficient of the location quotient for agriculture; Coef.5, Coefficient of industry location quotient; Coef.6, Coefficient of the location quotient for manufacturing; Coef.7, Coefficient quotient location of services; * Coefficient statistically significant at 5%, ** statistically significant coefficient 10%; GL, Degrees of freedom; LSDV, Method of variables with fixed effects dummies; D1 ... D5, five variables dummies corresponding to five different regions.**

Therefore, the data used and the results obtained in the estimations made, if we have conditional convergence, that will be in industry and all sectors.